\documentclass[twocolumn,prb,showpacs,superscriptaddress,amsmath,amssymb,natbib]{revtex4}

\usepackage[dvips]{epsfig}
\usepackage{graphicx}% Include figure files
\usepackage{dcolumn}% Align table columns on decimal point
\usepackage{bm}% bold math
\usepackage[english,german]{babel}
\usepackage{amsmath}
\usepackage{amssymb}
\usepackage[latin1]{inputenc}

\begin{document}

\title{Hidden complexities in the unfolding mechanism of a cytosine-rich DNA strand}

\author{Jens Smiatek}
\email{jens.smiatek@uni-muenster.de}
\affiliation{Institut f{\"u}r Physikalische Chemie, Universit{\"a}t
    M{\"u}nster, D-48149 M{\"u}nster, Germany} 
\author{Daniel Janssen-M{\"u}ller}
\affiliation{Institut f{\"u}r Physikalische Chemie, Universit{\"a}t
    M{\"u}nster, D-48149 M{\"u}nster, Germany}
\author{Rudolf Friedrich}
\affiliation{\selectlanguage{german}
  Institut f"ur Theoretische Physik, Universit{\"a}t M{\"u}nster, D-48149 M{\"u}nster, Germany
}
\author{Andreas Heuer}
\affiliation{Institut f{\"u}r Physikalische Chemie, Universit{\"a}t
    M{\"u}nster, D-48149 M{\"u}nster, Germany}

\selectlanguage{english}

\date{\today}

\begin{abstract} 
We investigate the unfolding pathway of a cytosine-rich DNA hairpin structure via Molecular Dynamics simulations. Our results indicate a hidden complexity present in the unfolding process. We show that this complex behavior arises due to non-stochastic contributions which have an important impact on the description of the dynamics in terms of collective variables.
\end{abstract}
\maketitle
\section{Introduction}
The study of macromolecular folding properties is of prior importance for modern nanotechnological research. The description of molecular dynamics in terms of generalized  variables allows to reduce the overwhelming 
atomistic motion to a few degrees of freedom. 
This is specifically of interest for protein engineering experiments as well as for the optimization of nanomolecular devices which are often built of DNA molecules \cite{1}.\\   
In the last decade the application of DNA as a novel construction material has grown enormously \cite{1}. 
One advantage is the effective control between the interactions of complementary strands due to the easy modification of 
the base sequences. In contrast to protein and RNA usage in nanoengineering, DNA additionally offers a simpler form and a better definition of  self-assembly processes.\\
Often used in this context are non Watson-crick structures like the DNA i-motif [2] which is a compact three-dimensional structure of cytosine-rich strands. It forms in an acidic environment where a 
cytosine-protonated-cytosine binding occurs which is able to stabilize a compact conformation \cite{3}. By systematic lowering and increasing the pH value it was found that the i-motif undergoes a reversible folding and 
unfolding process \cite{4} which can be even driven by a cyclic chemical reaction \cite{5,6}. It has been validated that this mechanism, which occurs on the timescale of  seconds, effectively performs work \cite{2}.\\ 
Technological applications are given by molecular nanomachines \cite{2,4}, switchable nanocontainers \cite{7} and building materials for logic gate devices \cite{8}. S
pecial attention has been spent on the properties of a grafted 
i-motif layer \cite{9}. It was found that the grafting density massively influences the folding properties of the i-motif. Hence it becomes clear that a detailed investigation of the unfolding respectively folding process 
is of prior interest for further optimization. \\
Due to the enormous complexity of the atomic motion, the progress of unfolding is often described in terms of simple collective variables \cite{10}. Collective variables offer the opportunity to systematically decrease the 
degrees of freedom such that the underlying dynamics can be described in a more unique fashion. Common examples used in the context of macromolecules are the end-to-end distance, dihedral components as well as any other 
variable that describes atomic motion. Thus it is obvious due to the above mentioned advantages that the usage of collective variables is a common method in theoretical as well as experimental studies.\\  
In the last years, several publications have reported hidden complexities \cite{11,12,13,14,15,16}in the description of unfolding processes in terms of collective variables. 
A nonlinear behavior concerning the slow relaxation dynamics of 
specific variables has been reported [12-16] which significantly reflects the hidden complexity involved in reaction coordinates. It has been discussed that the significant reduction of atomic motion in terms of 
collective variables may result in an inadequate description of the corresponding unfolding pathway \cite{12,13,14,15,16}.\\ 
Therefore the prediction of molecular motion and conformational changes in terms of collective variables for the optimization of nanomolecular machinery has to be proven carefully. It was suggested that the choice of 
inappropriate collective variables may also cause significant deviations in the corresponding free energy landscapes due to the presence of hysteresis effects \cite{17}.
In recent publications \cite{18,19} we have in detail discussed the corresponding stable conformations of a deprotonated i-motif as well as the possible transition pathways. The unfolding mechanism was mainly studied in 
terms of a principal component analysis \cite{18,19,20} with the corresponding eigenvectors. The eigenvectors are ranked due to the strength of the relative positional fluctuations of the atoms such that the lowest 
eigenvectors correspond to the main collective motion. It was shown that the key mechanism of unfolding can be described by the first two eigenvectors of the system where most of the important motion is 
involved \cite{18,19}.\\
We have shown \cite{18,19} that an unfolding mechanism between hairpin configurations and a fully unfolded structure is mainly oppressed due to energetic barriers and deep free energy minima. 
Additionally it was reported in Ref.~\cite{19} that high temperature simulations at 500 K are able to reconstruct the unfolding pathway in good agreement to 300 K. 
Thus the main unfolding pathways can be also analyzed at higher temperatures which 
allows a significant decrease of simulation times to observe rare events \cite{19}. It was also indicated that the global equilibrium structure of the hairpin configurations is shifted towards a fully unfolded strand 
at higher temperatures which was explained by a temperature-dependent entropy due to the interactions with the environment \cite{18}.\\ 
In this paper we want to study the transition mechanism between the hairpin configuration and a fully unfolded structure at high temperatures in detail. By distinct investigation of the corresponding distribution 
function in terms of their stochastic contributions, we are able to indicate an unexpected significant impact of the higher eigenvectors on the dynamics of the molecule in a specific region. 
More generally, our analysis reflects the strengths and weaknesses of a projection on a few collective variables and discusses probabilities for a recovery of the lost information which is disregarded in 
the applied reaction coordinates.\\ 
The paper is organized as follows. In the next section we present the numerical details of the simulations. The results are presented in section 3. We conclude with a brief summary in section 4.
\section{Numerical Details}
We have performed Molecular Dynamics simulations of a deprotonated i-motif in explicit TIP3P solvent at T=500 K by the GROMACS software package \cite{21}. The single DNA strand consists of 22 nucleic acid bases represented 
by the sequence 5'-CCC-TAA-CCC-TAA-CCC-TAA-CCC-T-3' where T, A and C denote thymine, adenine and cytosine. We modeled this structure which is nearly identical to the sequence used in \cite{4} 
by applying the molecular structure 
of the PDB entry 1ELN \cite{22} after changing uracil to thymine. The system contains 5495 TIP3P water molecules and 22 sodium ions (0.23 mol/L) to compensate the charging of the DNA. 
All interactions have been calculated by 
using the ffAMBER03 force field \cite{23}.\\ 
After energy minimization, the initial warm up phase of 1 ns has been performed by keeping the position of the i-motif restrained. 
The cubic simulation box with periodic boundary 
conditions has a length of 5.41 nm at each side. We applied a Nose-Hoover thermostat to keep the temperature constant. All bonds have been constrained by the LINCS algorithm. Electrostatics have been calculated by the 
PME algorithm \cite{24} and the time step was 2 fs. The free energy differences and barriers have been calculated at 300 K by the method and the simulations described in Refs.~\cite{18,19,25}. 
\section{Results}
The usage of high temperatures allows us to follow a complete unfolding process without the application of artificial methods like biasing potentials to overcome energetic barriers. It was shown in Ref.~\cite{19} that the 
unfolding pathway at 500 K largely agrees with the transitions found at 300 K due to the nearly identical motion of the first three eigenvectors as they have been estimated in the framework of the essential dynamics 
concept \cite{20}.\\ 
\begin{figure}[tbh]
  \includegraphics[width=0.5\textwidth]{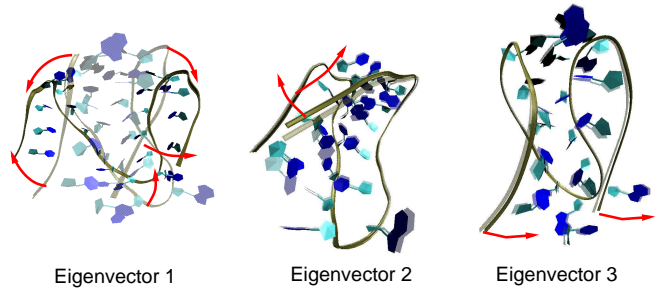}
  \caption{Illustration of the collective motion for the first three eigenvectors.}
        \label{fig1}
\end{figure}
For the study of the unfolding mechanism of the hairpin structures, we calculated the essential eigenvector subspace of the complete unfolding simulation as it has been in detail described in Refs. \cite{18,19}. 
We illustrate the first three eigenvectors as derived in the scope of a principal component analysis \cite{20} in Fig. ~\ref{fig1}. Eigenvector 1 represents an end-to-end relaxation of the end of the strands related to a 
broadening of the structure. The transition into a planar conformation is given by eigenvector 2 which describes the motion of the molecular ends perpendicular to the main loop. Eigenvector 3 includes torsional 
motion which results in a twisted configuration as denoted by the arrows.\\ 
In a recent publication \cite{19} it has been found that the dominant subspace of essential motion is formed by the first eight eigenvectors, whereas the combined first and the second eigenvector represent nearly 
66% of the overall atomistic fluctuations.
We have shown in Refs. \cite{18,19} that the most important unfolding pathway for a cytosine rich DNA strand at 300 K is given for the transition of hairpin structures to a fully unfolded configuration. The corresponding 
structures as well as the deprotonated unstable i-motif are presented in Fig.~\ref{fig2}. The arrows follow the unfolding pathway whereas the values for $\Delta F$ denote the free energy energy differences relative 
to the unstable 
structure. The highest free energy barrier found on each transition pathway is referenced by $\Delta$. 
By regarding each value, it becomes clear that the hairpin structures A and B are prevented by large barriers and steep minima from full unfolding into structure C.  The interconversion from A to B is achieved by a 
relaxation of the $5^{\prime}$ end.
In recent publications \cite{18,19} we have also indicated further pathways which allow a direct transformation from B to C as well as intermediate states resulting from a torsional motion. However, in this study we focus 
on the pathway shown in Fig.~\ref{fig2} due to the fact that it has been indicated as the most favorable unfolding pathway [18,19].\\ 
\begin{figure}[tbh]
  \includegraphics[width=0.5\textwidth]{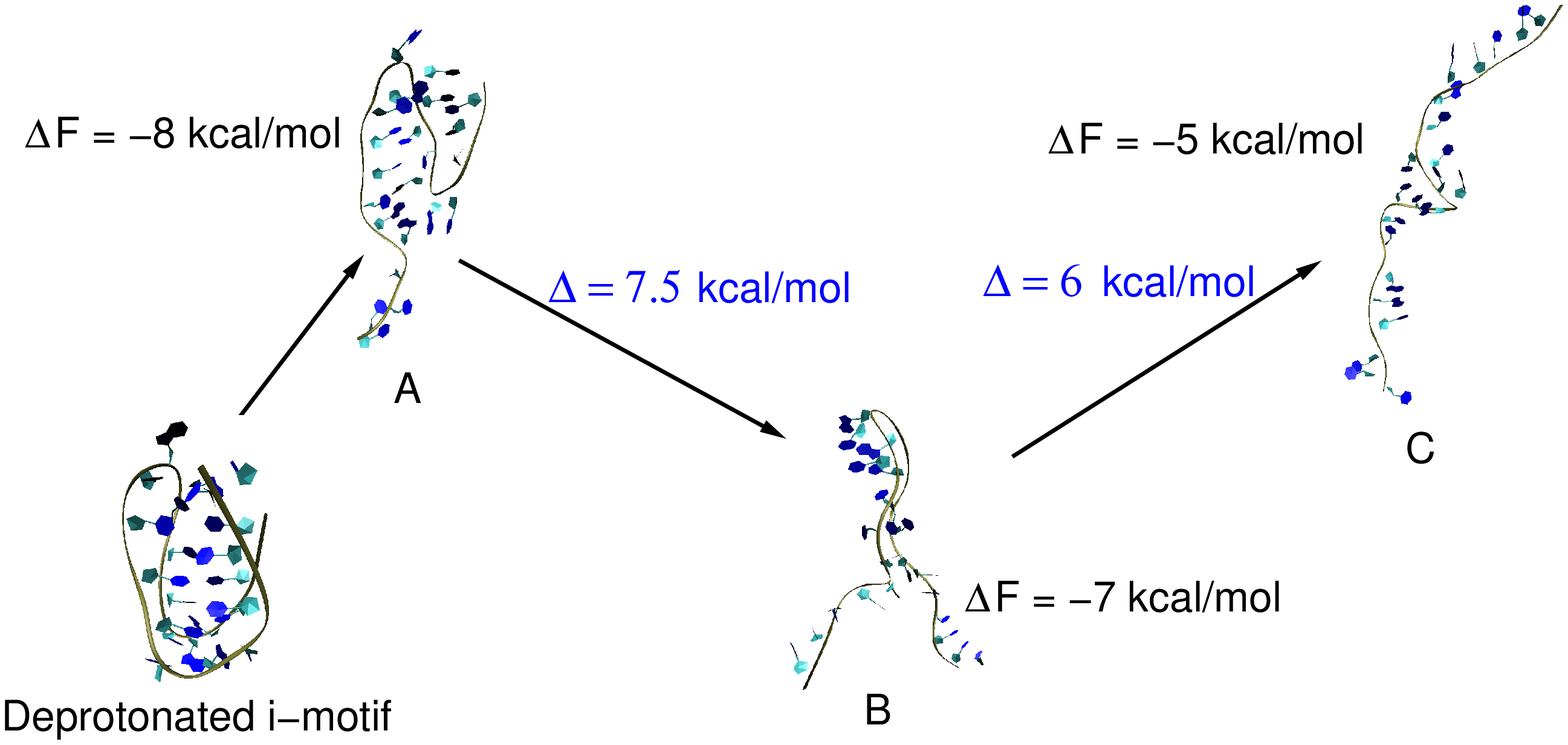}
\caption{Schematic illustration of the corresponding hairpin structures and the fully unfolded strand at 300 K. The arrows denote unfolding pathways and and the corresponding free energy differences relative to the 
starting structure (left) and the highest free energy barriers occurring on the corresponding pathways.}
        \label{fig2}
\end{figure}
The unfolding motion can be easily described by the first two eigenvectors if all other eigenvectors show a Gaussian distributed uncorrelated motion as it is often assumed when representing two-dimensional free energy 
landscapes \cite{10}. The validation for this choice is mainly given by the assumption of stochastic motion for the corresponding disregarded eigenvectors in absence of deterministic contributions \cite{14,15}. 
Specifically for essential dynamics it was assumed that the resulting stochastic distribution function for the non-essential and disregarded eigenvectors are Gaussian functions \cite{26,27,28}.\\ 
To investigate the third eigenvector as the highest disregarded collective variable on its stochastic contributions, the non-Gaussian parameterwith variance 
\begin{equation}
D= \frac{<\sigma^4>-3<\sigma^2>^2}{<\sigma^2>^2}
\end{equation}
yields as a quantitative measure for the deviation from a Gaussian distribution \cite{29}.\\
The results are presented in Fig.~\ref{fig3} where the deviations from a local Gaussian distribution function for eigenvector 3 in the landscape of eigenvector 1 and 2 are shown. 
Significant deviations can be observed in the 
range between -9 to -4 nm for eigenvector 1 and -5 to 25 nm for eigenvector 2. 
\begin{figure}[tbh]
  \includegraphics[width=0.5\textwidth]{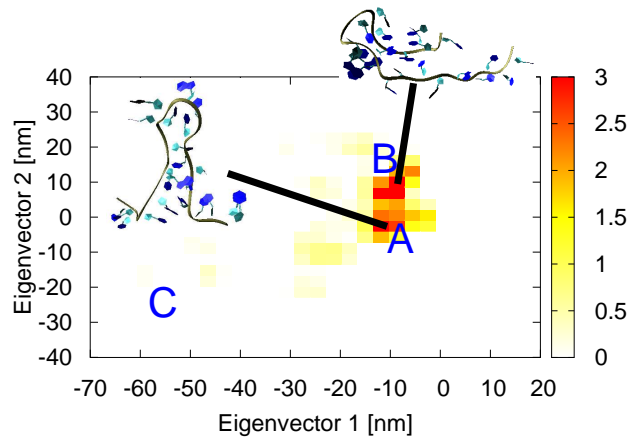}
\caption{Non Gaussian parameter D of the third eigenvector as determined in the landscape of the first and the second eigenvector. Each bin consists of minimally 50 points and the corresponding energetic minima 
are denoted by letters in agreement to Fig.~\ref{fig2}.}
        \label{fig3}
\end{figure}
It is obvious that significant deviations from a Gaussian distribution function for eigenvector 3 can be mainly indicated at the region in close vicinity of the hairpin structures. Thus it can be assumed that 
eigenvector 3 has a significant impact on the unfolding motion of the hairpin configurations (A,B) into a fully unfolded strand (C). It can be concluded that the two-dimensional representation of the unfolding pathway 
in terms of eigenvector 1 and 2 is not sufficient due to the neglect of torsional motion as given by eigenvector 3.\\
In order to get a more detailed understanding of the underlying complexity, it comes out that eigenvector 3 shows significant non-stochastic motions as shown in Fig.~\ref{fig4} 
\begin{figure}[tbh]
  \includegraphics[width=0.5\textwidth]{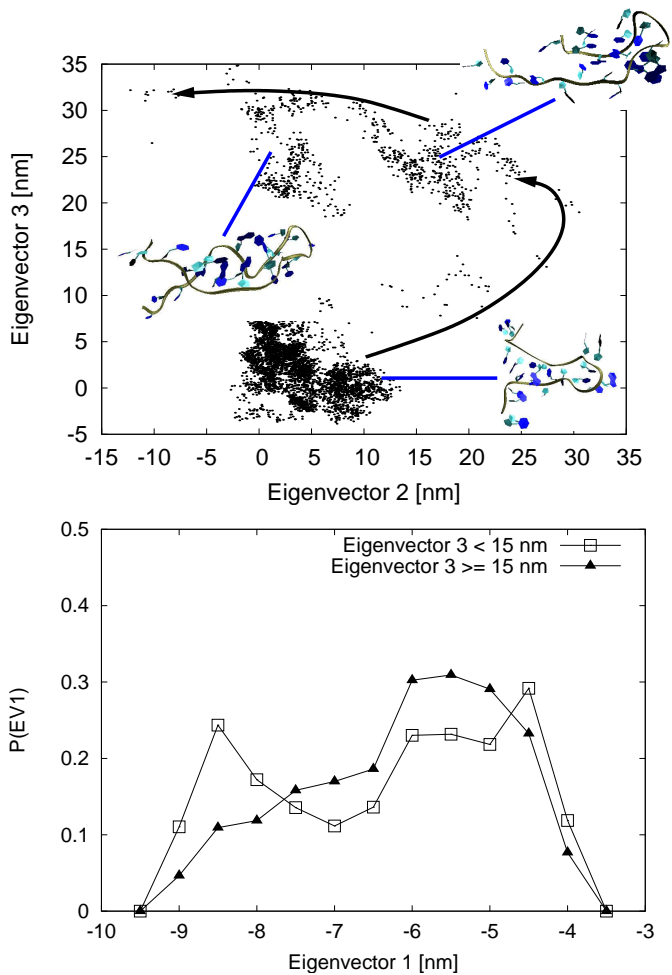}
\caption{Values for the eigenvector 3 in correlation with eigenvector 2 for the range -9 to -4 of the first eigenvector. The arrows denote the evolution in time. 
The clustered points can be related to a torsional motion of the hairpin structure. The corresponding distribution function for the first eigenvector is presented at the bottom.}
        \label{fig4}
\end{figure}
The corresponding structures are 
following an evolution in time as denoted by the arrows. By detailed regard of the unfolding progress into the fully unfolded strand it becomes obvious that a transition mainly occurs due to the contribution of the 
torsional motion of eigenvector 3. Hence the hairpin structures are enforced to twist for the breakage of the intramolecular hydrogen bonds \cite{18,19} which stabilize the hairpin structures 
whose impact finally results in a random coil structure (c. f. Fig.~\ref{fig2}).\\ 
Furthermore it has to be noticed that eigenvector 1 is nearly independent on the motion of eigenvector 3 as it is also true for eigenvector 2. This can be seen by analyzing the corresponding distribution function 
as it is shown in the bottom of Fig.~\ref{fig4}. It is obvious that the mean values for eigenvector 1 are nearly comparable as they are -6.33 nm (Eigenvector 3 $<$ 15 nm) and -6.11 (Eigenvector 3 $>$ 15 nm). Hence the 
fluctuations of eigenvector 3 are not correlated to eigenvector 1 due to the fact that the distribution functions are overlapping. This is also in particular true by regarding the motion for eigenvector 2 
which starts at 5 nm, extends to 25 nm and finally returns back to 5 nm. Thus the two structures at an eigenvector 2 value of 5 nm can only be separated by the application of the third eigenvector.  
These results again emphasize the importance of the third eigenvector for the unfolding motion. It can be concluded that a separation of several conformations at a specific region can only be achieved in a 
higher dimensional landscape.\\
In addition we have investigated the non Gaussian parameter $D$ for the eigenvectors 4 to 12 in the specific region as noticed above. The results for $D$ are presented in Fig.~\ref{fig5}.  
\begin{figure}[tbh]
  \includegraphics[width=0.5\textwidth]{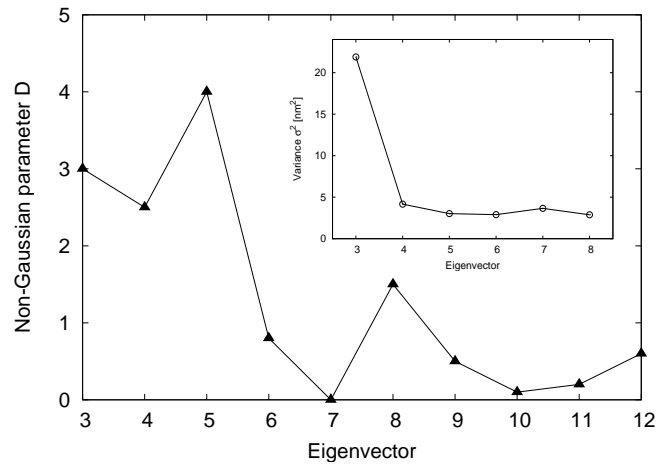}
\caption{Non Gaussian parameter $D$ in the specific region for eigenvector 1 with -9 to -4 nm and eigenvector 2 with 0 to 20 nm for eigenvectors 3 to 12. Inset: Variances for the eigenvectors in the specific region.}
        \label{fig5}
\end{figure}
It comes out that eigenvectors 
4 and 5 also display significant non Gaussian defects. This means that the unfolding described in this region is mainly driven by non-stochastic contributions of the first five eigenvectors. However, as the 
inset of Fig.~\ref{fig5} also shows, the variances for these distributions are small compared to eigenvector 3. This indicates that the local fluctuations for eigenvectors larger than 
3 are smaller compared to lower eigenvectors. Hence the importance of this motion is diminished in the light of a large collective motion.
\section{Summary}
We have investigated the unfolding mechanism of stable hairpin structures into a fully unfolded structure in detail. Our results indicate a significant contribution of eigenvectors 3 to 5 to the unfolding motion in 
combination with eigenvector 1 and 2.  Hence we conclude that the motion of these eigenvectors which directly belong to the unfolding of the hairpin configurations is largely dominated by non-stochastic contributions. 
We have therefore shown that a complete description of the unfolding motion of a cytosine-rich DNA hairpin configuration has to include the torsional motion of the strand ends as it is expressed in eigenvector 3 and 
higher eigenvectors close to the configurational vicinity of the energetic minima.\\
A general method has been presented for the study of deviations from a local Gaussian distribution which result in explicit non-stochastic motion. The method is easy to generalize for various collective variables. 
For a quantitative estimate we have introduced the non Gaussian parameter $D$. To the best of our knowledge, we have not found any related work which uses this characteristic value in the context of hidden 
complexities in biomolecular unfolding pathways. In general it can be concluded, that deviations from a Gaussian distribution result in a non-Markovian behavior of the dynamics \cite{26,27,28}. 
In detail, we have shown that 
the knowledge of eigenvector 1 and 2 is not enough for an explicit prediction of the future dynamics of the system.\\
Summarizing the results, we have validated that the unfolding process of hairpin structures is a combined approach of relaxational motion as given for eigenvectors 1 and 2 and the torsional motion of eigenvector 3. 
For a systematic description of the unfolding pathway the end-to-end relaxation, orthogonal movements of the strand ends to the planar loop as well as torsional motion has to be directly taken into account. 
On the other hand by disregarding these movements by inadequately chosen collective variables, important information is lost such that the process of unfolding remains in detail incomplete.
This sheds a new light on the importance of the unfolding pathway in a crowded environment as given for grafted DNA i-motifs. It can be assumed that in a high density environment, several aspects of this study 
have to be modified due to the strict investigation of single-molecular motion. However, it can be also concluded that for a complete unfolding transition, even in a crowded environment, the torsional motion 
is essential. Hence, by an oppression of this motion due to steric reasons a complete unfolding can be hardly observed.\\  
In addition to energetic reasons \cite{10,11} we have therefore found a second explanation for the stability of the hairpin structures at neutral pH values due to an enhanced complexity of the unfolding pathway. 
Our results have therefore shown, that the choice of appropriate collective variables is not obvious. We have presented methods for a distinct investigation of the chosen collective variables in terms of 
hidden complexity to tackle this challenge.\\
Nevertheless, the problem of an adequate description of unfolding progress in terms of generalized variables is a remaining problem both for theorists as well as experimentalists.
\section{Acknowledgements}
We thank the Deutsche Forschungsgemeinschaft (DFG) through the Transregional collaborative research center TRR 61 for financial funding.

\end{document}